\documentclass{elsart}
\usepackage{psfig}

\newcommand{\BE}{\begin{eqnarray}}
\newcommand{\EE}{\end{eqnarray}}
\newcommand{\BEn}{\begin{eqnarray*}}
\newcommand{\EEn}{\end{eqnarray*}}
\newcommand{\barr}{\begin{array}}
\newcommand{\earr}{\end{array}}

\newcommand{\bit}{\begin{itemize}}      
\newcommand{\eit}{\end{itemize}}
\newcommand{\bc}{\begin{center}}
\newcommand{\ec}{\end{center}}
\newcommand{\ben}{\begin{enumerate}}    
\newcommand{\een}{\end{enumerate}}

\newcommand{\be}{\begin{equation}\FL}
\newcommand{\ee}{\end{equation}}
\newcommand{\beas}{\begin{eqnarray*}}
\newcommand{\eeas}{\end{eqnarray*}}
\newcommand{\bea}{\begin{eqnarray}\FL}
\newcommand{\eea}{\end{eqnarray}}

\newcommand{\eps}{\epsilon}

\def\bit{\hbox{\it I\hskip -2pt  B}}

\begin{document}

\begin{frontmatter}
\title{Minority Games and stylized facts}
\author{Damien Challet${}^{(1)}$, Matteo Marsili${}^{(2)}$ and 
Yi-Cheng Zhang${}^{(3)}$} 
\address{${}^{(1)}$ Theoretical Physics, 
Oxford University, 1 Keble Road, Oxford OX1 3NP, United Kingdom\\
${}^{(2)}$ Istituto Nazionale per la Fisica della Materia ({\it INFM}),
Unit{\'a} di Trieste-SISSA, I-34014 Trieste, Italy\\
${}^{(3)}$
Institut de Physique Th{\'e}orique, Universit{\'e} de Fribourg, 
Perolles CH-1700, Switzerland}
\date{\today}

\begin{abstract}
The Minority Game is a generic model of competing adaptive agents, which is often believed to be a model of financial markets. We discuss to which extend this is a reasonable statement, and present  minimal modifications that make this model reproduce stylized facts. The resulting model shows that without speculators, prices follow random walks, and that stylized facts disappear if enough speculators take into account their market impact.
\end{abstract}
\begin{keyword}
Minority Game, financial markets, stylized facts
\end{keyword}

\end{frontmatter}
\section{Introduction}
The Minority Game \cite{CZ97} was introduced a few years ago as 
a physicists' simplification of the famous Arthur's  El Farol's Bar problem \cite{Arthur}. Since then it has attracted much attention from the econophysics community \cite{web}. The MG was not initially thought as a model of financial market, but rather as a generic model of competing adaptive agents in economy. Nevertheless, a majority of papers on the MG are motivated by study of financial markets. It is therefore worth investigating why and to which 
extend this is justified. 

In this game $N$ agents have to select one choice between two at each time step, and those who are in minority are rewarded. They do not act randomly, but rather inductively. This is achieved by giving to all agents their own set of $S$ strategies, or theories of the world, which predict a winning action for all $P$ possible states of the words. The agents are inductive in the sense that they assign a score to each of their strategies which reflect their perceived performance and use  a given  time the one with the highest score. The strategies' sets of all agents are fixed before the beginning of the game, hence, play the role a quenched disorder. If agents have no memory ($P=1$), there is no frustration in the physical sense,  and an exact solution is straightforward \cite{MC00}. If $P>1$, frustration arises because agents cannot optimize their behavior simultaneously for all states of the world. In that case, the spin glass nature of the MG is  is revealed by a fruitful mathematical formalism \cite{CM99,CMZe99,MCZe99}; spin-glass techniques yield solve the model \cite{CMZe99,MCZe99,CMZ99,Coolen}.

The minority mechanism has three fundamental consequences:

\begin{enumerate}
\item Competition for limited resources: not all agents can win at the same time\footnote{Note that competition only causes a {\em psychological} frustration, but no physical frustration: the latter is due to the memory of agents, as stated above.}. 
\item There is no good behavior: a behavior is good only with respect to  others' behavior
\item A good behavior may become bad when others' behavior changes.
\end{enumerate}

Agents' inductive behavior complete the definition of the game:
\begin{enumerate}
\item[(4)] Adaptive agents try to predict next winning choice, which is determined only by their own choices.
\end{enumerate}

\section{MG and financial markets}

In a metaphorical way, this sounds like a financial market\footnote{The question why financial markets can be modeled by a minority mechanism is discussed in ref \cite{Mproc}.}. At this point, there are however several 
characteristics that the MG does not share with financial markets. 

\subsection{Producers}

The first problem is that 
the MG is a negative sum game, hence, it is unclear why speculators would be willing to play such games. Indeed, in the basic MG, agents are forced to play at each time step; one expects that no agent would remain after some while in the game if they were allowed not to play. This amounts to ask why speculators are interested in real markets.

The money does not come from the speculators themselves; this suggests that there are other types of agents, in particuler agents called producers in ref \cite{ZMEM} who are not interested in making money inside markets, but who use the market for exchanging goods. They introduce some predictable patterns in the market which speculators exploit for their own profit. The producers are much less adaptive than the speculators; in the MG context, this is reflected by giving significantly less strategies to the producers than to the speculators; as in ref \cite{CMZ99}, we consider $N_p$ producers (non-adaptive\footnote{i.e. with one strategy only.} agents) and $N_s$ speculators (standard inductive agents). The resulting model already allows one for studying the interplay between the information content $H$ left by producers' behavior, and the gain of the speculators. In addition, it is also exactly solvable by spin-glass techniques.

\subsection{A realistic grand canonical mechanism}

\begin{figure}
\centerline{\psfig{file=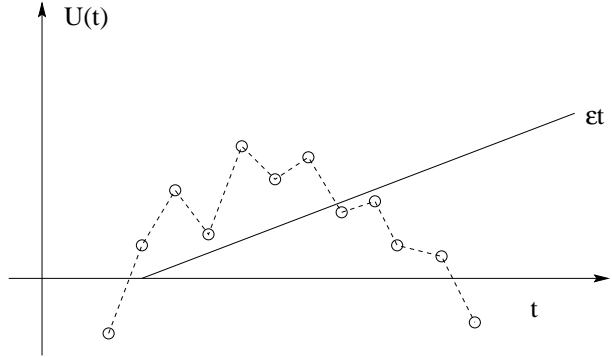,width=8cm}}
\caption{Graphical illustration of the proposed grand canonical mechanism. $U(t)$ is the score of the best strategy of the considered agent.}
\label{grcanmech}
\end{figure}

Now that agents have a good reason to enter into the market, they still need a criterion which should tell them when to enter and when to withdraw. Let us review carefully agents' behavior.
For an agent of the basic MG, being inductive means  {\em comparing} the performance of all her strategies and nothing more. In other words, only the {\em relative} strategies' value is considered by agents, whereas actual value of scores does not matter; consequently, an agent is not worried by her real gain\footnote{This makes sense as long as she is forced to play.}. But having a positive or negative real gain is of great relevance in reality. Therefore, it seems reasonable to consider the following grand canonical mechanism: at a given time, an agent plays if she has at least one strategy with a positive score\footnote{Mathematically, if $U_{i}^{\rm max}(t)$ be the score of her best strategy: the criterion is simply play if $U_{i}^{\rm max}(t)>0$.} \cite{J99,CMZ99,BouchMG}. However, this mechanism is problematic for two reasons. First, if there are no producers, hence no incentive for the speculators to play, one ends with about $75\%$ of speculators in the market if $N_s$ is large enough\footnote{Except if $N<N_c=P/\alpha_c$ \cite{Bouch}.}. A more subtle but even more important, problem is the fact that an agent who enters into market at time $t$ and withdraw from it at time $t+\Delta t$ is sure to suffer a loss of at least $U(t)-U(t+\Delta t)$ where $U(t+\Delta t)<0$; indeed, the increase of score --- a possible gain --- between $t-1$ and $t$ is {\em virtual}, since the agent is not in the market, whereas the loss is real (see figure \ref{grcanmech}).

Consequently, there is a need of another grand canonical mechanism\footnote{The mechanism we proposed in \cite{CMZ01} is essentially similar to the one of ref \cite{J00}; however, in our version, the scores are kept since the beginning of the game, not only during a small past time-window. The latter assumption is probably more realistic, but is not needed for our purpose.}. The idea is to give a benchmark to agents such that they only stay in the market if they perform well {\em in the market} and not only outside from it; as a consequence, the more time they can spend in the market, the more successful they are. The benchmark we propose is simply that an agent only plays if she has at least one strategy with a score higher than $\eps t/P$ ($t/P$ is the system size's independent time; see \cite{MCZe99}).  Note that it is clear from figure \ref{grcanmech} that the benchmark will make agents withdraw quicker from the market. 

The $\eps$ parameter consists of two parts. The first one is a ``common sense'' factor $C$, which remedies the sure loss problem of the $\eps=0$ case; the second one can be interpreted as the interest rate $I$ of a risk-free account, hence $\eps=C+I$, that is, even if the interest rate is zero, one still has to consider $\eps>0$. Strikingly, as soon as $\eps>0$, there is a phase transition of first order for $n_s=N_s/P>n_s^*$ and the average number of speculators inside the market is proportional to the amount of information left by the producers\footnote{Note that here $n_s=N_s/P=1/\alpha$, where $\alpha$ is the usual control parameter \cite{Savit}.}. In addition, in this region,  a whole set of stylized facts \cite{MantegnaStanley} arises: clustered volatility, power-law tails of returns' and volume's distribution can be obtained. Note also that this region is {\em marginally} efficient if $\eps>0$; when the number of speculators increases,  the market becomes more efficient. Since producers' contribution to the price dynamics is binomial, the price follows a random walk without the speculators, and if there are enough speculators, stylized facts arises\footnote{The above behavior {\em crucially} depends on the price-taking behavior of a large fraction of the speculators. If there are enough speculators who account for their impact, all stylized facts disappears.}.

\begin{figure}
\centerline{\psfig{file=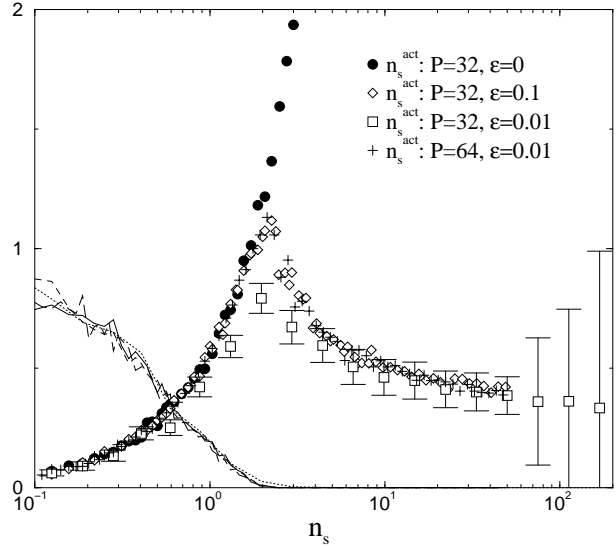,width=8cm}}
\caption{Predictability per agent $H/(N_p+N_s)$ (lines) and average number of
active speculators $n_s^{\rm act}$ (symbols) as a function of $n_s=N_s/P$
for $N_p=P$ and some values of $\epsilon$ and $P$. For $n_s>n_s^*$, $n_s^{\rm act}$
has a discontinuous behavior as $\epsilon\to 0$. In addition,  as soon as $\eps$, $n_{s}^{\rm act}$ does not depend on  $\eps$ and tends to a constant value for $n_s\to \infty$}
\label{Hnsact}
\end{figure}

Therefore, the combination of the presence of producers and the proposed grand canonical mechanism not only answer to the questions of why and how should speculators participate to markets, but also reproduces some markets' characteristics. Most importantly in a MG context, this combination is still exactly solvable \cite{CMRZ}.

\subsection{Evolving capital -- re-investment}
This extension has been considered in \cite{J00,FarmerJoshi,CCMZ00}. Even if evolving capitals are not needed in order to reproduce stylized facts in MG-like models, it has a particular economic relevance, so that it makes sense to complete the above models with this feature. In this case, the presence of producers is needed, else all speculators are ruined after some while; even if agents are forced to play, near the critical point stylized facts also arise; the resulting model is also exactly solvable \cite{CCMZ00}, although much more difficult to tackle analytically.

\section{Conclusions}

Very minimal modifications are need in order to obtain stylized facts in MG-like models. What we believe to be the minimal modification to the standard MG is the combination of producers and speculators playing or not according to their  benchmark. A more realistic and still exactly solvable model of markets is a MG with producers, benchmark and evolving capitals, but the latter are not needed in order to produce stylized facts.

\end{document}